# Photodetection in p–n junctions formed by electrolyte-gated transistors of two-dimensional crystals

Daichi Kozawa[1,2,*], Jiang Pu[3], Ryo Shimizu[1], Shota Kimura[1], Ming-Hui Chiu[4], Keiichiro Matsuki[3], Yoshifumi Wada[1], Tomo Sakanoue[1,2], Yoshihiro Iwasa[5,6], Lain-Jong Li[4], Taishi Takenobu[1,2,3,7,*]

[1] Department of Applied Physics, Waseda University, Shinjuku, Tokyo 169-8555, Japan
[2] School of Engineering, Nagoya University, Furo-cho, Chikusa-ku, Nagoya, 464-8603, Japan
[3] Department of Advanced Science and Engineering, Waseda University, Shinjuku, Tokyo 169-8555, Japan
[4] Physical Science and Engineering Division, King Abdullah University of Science and Technology, Thuwal 23955-6900, Kingdom of Saudi Arabia
[5] Quantum-Phase Electronics Center and Department of Applied Physics, The University of Tokyo, Tokyo 113-8656, Japan
[6] RIKEN Center for Emergent Matter Science, Wako 351-0198, Japan
[7] Kagami Memorial Laboratory, Waseda University, Shinjuku, Tokyo 169-0051, Japan
[*]dkozawa@mit.edu
[*]takenobu@nuap.nagoya-u.ac.jp

Transition metal dichalcogenide (TMDC) monolayers have attracted much attention due to their strong light absorption and excellent electronic properties. These advantages make this type of two-dimensional crystal a promising one for optoelectronic device applications. In the case of photoelectric conversion devices such as photodetectors and photovoltaic cells, p–n junctions are one of the most important devices. Here, we demonstrate photodetection with $WSe_2$ monolayer films. We prepare the electrolyte-gated ambipolar transistors and electrostatic p–n junctions are formed by the electrolyte-gating technique at 270 K. These p-n junctions are cooled down to fix the ion motion (and p-n junctions) and we observed the reasonable photocurrent spectra without the external bias, indicating the formation of p-n junctions. Very interestingly, two-terminal devices exhibit higher photoresponsivity than that of three-terminal ones, suggesting the formation of highly balanced anion and cation layers. The maximum photoresponsivity reaches 5 mA/W in resonance with the first excitonic peak. Our technique provides important evidence for optoelectronics in atomically thin crystals.

The p–n junction is a key component of optoelectronic devices based on two-dimensional (2D) crystals, such as p–n diodes,[1] light-emitting diodes,[2-4] photodetectors,[3-5] and photovoltaic cells.[6] In particular case of photoelectric conversion devices based on p–n diodes, photogenerated electron–hole pairs (or excitons) are readily separated under built-in potential in a material depletion region and migrate toward the electrodes. The p–n photodiode is one of the most important tools for investigating the fundamental optoelectronic properties of materials. In recent years, p–n junctions have been formed in transition metal dichalcogenide (TMDC) monolayers in various ways. One is in ambipolar field-effect transistors[7,8] based on solid dielectric layers ($SiO_2$) with independent two-input local gates.[2,4,6] Others include the use of spatially selective chemical doping[9] and lateral p–n junctions. The significant point here is that these p–n junctions enable electroluminescence and photoelectric conversion, which offers the possibility of photoresponsivity for future applications.[2,4,6,10] However, the formation of p–n junctions in TMDCs is limited to certain mechanisms such as chemical doping[9] and electric field doping on split gates.[2,4,6] In particular, because the Fermi level of $MoS_2$ is highly doped in favor of an n-type semiconductor, high-density charge doping is required to obtain p–n junction.[11]

In order to establish a universal technique to achieve photodetection with a lateral p–n junction of semiconducting TMDC, we develop a photodetection technique based on an electrolyte-gated monolayer TMDC transistor. This type of transistor is gated by inducing an electric double layer, and is referred as an electric double-layer transistor (EDLT).[12,13-15] Its high gate efficiency enables an extremely high carrier density ($>10^{14}$ carriers/$cm^2$), ambipolar operation,[7,13,16] and even field-induced superconductivity in TMDCs.[17] Electrostatic p–n junctions in TMDCs have been demonstrated in ambipolar EDLTs via the freezing-while-gating technique with no bias application.[14,15] This involves freezing the biased electrolyte to restrict the movement of ions, and then resulting in built-in electric fields with no external bias. In addition, this device structure is suitable for detailed photocurrent spectroscopy because the high carrier density due to electric double layers can effectively reduce the effect of trapping states in the semiconductor/electrolyte interfaces than those of semiconductor/solid dielectrics (e.g. $SiO_2$).[15,18] Consequently, the EDLT technique is promising for p–n photodiodes with high built-in electric fields.

Other important photodetectors based on TMDC monolayers are phototransistors.[19,20,21] Although photodiode consists of p-n junction at the interface of p- and n-type semiconductors, a key component of phototransistor is mainly Schottky junctions formed at the interface of semiconductor and metal electrodes. In phototransistors, illumination of light induces photogenerated carriers to be photoconductive, which can be amplified by carrier accumulation under applied gate voltage. Importantly, the phototransistors with excellent performances have already been reported.[20,21,22] Because this idea is also applicable to the EDLT technique, we will argue whether our devices are p-n photodiodes or phototransistors. It should be noted that the critical difference between p-n photodiodes and phototransistors is the external bias dependence and only p-n photodiodes can detect photocurrent without the external bias, which is preferable to realize lower power consumption photodetectors.

Here, we demonstrate a p–n photodiode without an external bias based on an electrolyte-gated monolayer $WSe_2$ transistor, in which monolayer $WSe_2$ is ready for ambipolar operation and suitable for the purposes of demonstration. We report the results of photocurrent spectroscopy to examine the photoelectric performance of the p–n photodiode, which is a crucial factor in understanding the fundamental optoelectronic properties. The resulting photocurrent spectra are valid and apparent excitonic features are observed. Our work shows that an electrolyte-gated p–n photodetector is a useful tool for investigating optoelectronic properties in monolayer TMDCs.

Monolayer $WSe_2$ was synthesized on a sapphire substrate using chemical vapor deposition (CVD) following our previous work.[23] This centimeter-scale film enables simple optical and electrical measurements, and is suitable for integration into optoelectronic devices. The sample was optically characterized by absorption, Raman, and PL spectroscopy and verified to be monolayer (Fig. S1 in the supplementary material). Absorption spectroscopy was performed in a UV-Vis-NIR spectrophotometer (Varian, Cary 5000), and Raman and PL spectroscopy were conducted in a spectrophotometer (JASCO, NRS-5100).

As an electrolyte, we selected an ion gel, which consisted of an ionic liquid ([EMIM][TFSI], 1-ethyl-3-methylimidazolium bis (trifluoromethylsulfonyl) imide) and polymer (PS-PMMA-PS, poly (styrene-b- methylmethacrylate-b-styrene)). The ethyl propionate solution of ion gel was spin-coated on the device. The electrolyte-gated transistors were characterized using a semiconductor parameter analyzer (Agilent Technologies, Inc., E5270) in a shield probe station inside an $N_2$-filled glove box at room temperature. For the light illumination experiments, we transferred these devices to a temperature controllable probe station with an observation glass window (Lake Shore Cryotronics, TTPX) and measurements were carried out using another semiconductor parameter analyzer (Agilent, B1500A) under high vacuum ($<10^{-3}$ Pa).

For the electrolyte-gated ambipolar transistors, using monolayer CVD-grown $WSe_2$ as the channel material, we fabricated three devices to investigate the reproducibility of our results (Device I, Device II and Device III). A picture and schematic image of Device I are shown in Fig. 1(a). We designed a large active channel area with a channel length of 400 μm and width of 1.2 mm to prepare enough space for the light illumination. Figure 1(b) shows the transfer characteristics of the Device I (also see Fig. S2 in the supplementary material). Similar to previous reports,[2,23,24] we observed ambipolar transistor behavior in our monolayer $WSe_2$ EDLTs. Using the specific capacitance measured via impedance spectroscopy (~5 μF/cm$^2$), we calculated relatively high hole and electron mobilities of 65 and 2.3 cm$^2$/Vs, respectively (Fig. S3 in the supplementary material). We also examined the p- and n-channel output characteristics (Figs. 1(c) and 1(d)). The linear increase and subsequent saturation of the current signify ohmic-like contacts in the metal electrodes/WSe$_2$ monolayer interfaces, suggesting a negligible Schottky barrier to both hole and electron injections.

Because we observed the ambipolar transistor behavior in our monolayer WSe$_2$ EDLTs, we tried to form p-n junctions electrostatically, following the reported method.[15] As schematically illustrated in Figs. 2(a) and 2(b), in ambipolar EDLTs, one can accumulate both electrons and holes simultaneously inside the channel, particularly in ambipolar regions, where drain voltage, $V_D$, is larger than gate voltage, $V_G$. For the p-n junction formation, we applied positive $V_G$ and measured the output characteristics at 270 K, as shown in Fig. 2(c). We selected 270 K to reduce the possibility of electrochemical reactions and this temperature is above the glass transition temperature of the ionic liquid. Similar to Fig. 1(d), ideal ohmic-like contacts and a reasonable saturation behavior were observed, indicating a weak effect of the Schottky barrier for electron injections. However, in the range of higher $V_D$ than that of Fig. 1(d) ($V_D > 2.1$ V), we observed the nonlinear amplification of drain current. Very importantly, it is already well established that the formation of p-n junction is verified by this sharp increase of drain current, as reported in Ref. [25] If $V_D$ is markedly below the applied $V_G$, the accumulation layer is formed by consisting of only electrons (Fig. 2(a)). However, if $V_D$-$V_G$ exceeds a specific threshold voltage (Fig. S2 in the supplementary material), the polarity of the effective gate voltage at the drain electrode reverses to accumulate holes, resulting in the formation of electrostatic p-n junctions, as schematically shown in the Fig. 2(b).

After determining the signature of the electrostatic p–n junctions at a drain voltage of 3.2 V and a gate voltage of 3.0 V, EDLT was cooled down to 160 K under constant drain/gate biases of $V_D = 3.2$ V and $V_G = 3.0$ V, respectively, with a rate of 0.5–1.0 K/min to restrict the movement of ions. Because 160 K is below the glass transition temperature of the ionic liquid, we can expect the formation of stable p–n junction.[15] Therefore, in order to confirm the formation of stable p–n junction, we released both drain/gate biases and measured the output characteristics without gate bias, in which we obtained obvious diode behavior with a rectification ratio of 400 at ±3 V (Fig.2(d) in linear and Fig. S4 in the supplementary material in log scale).

The measured current/voltage characteristics enabled determination of the band alignment of the p–n junction as a consequence of the ambipolar charge accumulation. Under high-density ambipolar carrier accumulation (Fig. 2(b)), the band diagram follows the alignment shown in Fig. 2(e). Because the specific capacitance of the electric double layer is very large (~5 µF/cm$^2$), the carrier density of doped region is typically $10^{13} – 10^{14}$ /cm$^2$. Therefore, the p- and n-type doped regions are close to degenerated semiconductors, while the depletion region has less carrier density. As a result, voltage drop in the p- and n-type regions is negligible due to their low resistance, and almost all drain voltage can be concentrated on the highly resistive depletion region. Very interestingly, these arguments also suggest the band alignment of the p–n junction at low temperature (Fig. 2(f)), which corresponds to the situation of Fig. 2(d). When EDLT was cooled down to 160 K under constant drain/gate biases of $V_D$ = 3.2 V and $V_G$ = 3.0 V, respectively, the band alignment of the p–n junction is still similar to Fig. 2(e). However, once we release both drain/gate biases at 160 K, the Fermi levels of p- and n-type regions will be aligned by mobile holes and electrons without movements of anions and cations, resulting in the flip of built-in potential and the band alignment in Fig. 2(f). In this Fermi levels alignment, the anions and cations correspond to the space charges, and therefore, due to the formation of p-n junctions without the external bias, we examine the photodetection capability of the device.

After achieved bias-free p–n junctions at 160 K, white light-emitting diodes (LEDs) light irradiation tests were performed without drain voltages to investigate the photodetection capability of the device (Fig. 3(a)). The luminescence spectrum of the LED is shown in Fig. 3(b) and indicates sufficient energy overlap with the absorption spectrum of the monolayer $WSe_2$ films. $WSe_2$ monolayers were irradiated using the LED light through the ion gel films. As indicated in Fig. 3(c), a dependence between the source/drain current and time/incident light power was observed, although no bias was applied to the source/drain/gate electrodes. The power of the light source was measured by an optical power meter (Newport Corporation 1930-C). The time-dependent response was calibrated using a Si photodiode (Hamamatsu Photonics K.K. S2281-04) with a bias of -1 V. We define photocurrent here as the difference between the illuminated and dark currents. The response rise time (from 10% to 90% of the maximum photocurrent when switching the light from OFF to ON) and fall time (from 90% to 10% of the maximum photocurrent when switching the light from ON to OFF) were much faster than the minimum time resolution associated with the measurement setup (100 ms), which is consistent with previous results in CVD-grown $MoS_2$ films.[21,26]

To confirm the reproducibility of the photodetection capability in $WSe_2$ EDLTs, we fabricated another device (Device II), in which device structure is similar to Device I. As shown in Fig. S5 in the supplementary material, Device II reproduce the results of Device I very well. Moreover, because the circular polarization was reported in electroluminescence spectra of similar p-n junctions in single crystal $MoS_2$ and $MoSe_2$, we also investigated the chirality dependence of the incident light in the photocurrent, resulting in no obvious chirality dependence at 77 K (Fig. S6 in the supplementary material). However, these results do not fully deny the chirality dependence in our photodetectors completely, because our CVD-grown $MoS_2$ films do not also show the chirality dependence in photoluminescence spectra. In addition, the hysteresis observed in the transfer curve shown in Fig. 1(b) does not affect photoresponses because this hysteresis originates from the slow response of ion migrations to the sweep rate of $V_g$ and these ions are fixed in these temperatures.

To further investigate the photo detecting properties of this photodiode, we conducted excitation-energy-dependent photoresponsivity measurements. For these, we used a comb-like configuration for the source and drain electrodes (Ni/Au electrodes 3/80 nm) to maximize the source/drain current (Device III, see Fig. S7(a) in the supplementary material). In the photoresponsivity measurements, the devices were illuminated by either an LED lamp (Olympus, SZ-LW61) or a Xe lamp (Asahi Spectra, Lax-Cute) combined with a monochromator (Asahi Spectra, CMS-100) under a back-scattering geometry in vacuum (<$10^{-3}$ Pa). The linewidth of optical excitation was 7–33 meV depending on excitation wavelength (Fig. S8 in the supplementary material). The beam illuminating the device was collected with a rectangular spot size of 3.5 mm × 9.7 mm. Its power range was 30–75 μW, over which the photocurrent had a linear dependence on excitation power (Fig. S9 in the supplementary material). The photoresponsivity spectrum was determined as the photocurrent divided by the excitation power as a function of excitation photon energy. The black dots in Fig. 4 represent the photoresponsivity of Device III without the external bias at 85 K, in which we applied $V_G = 1.6$ V and $V_D = 3.0$ V during freezing. The obtained photoresponsivity spectrum followed the shape of the absorption spectrum and exhibited four peaks corresponding to the excitonic transitions of A, B, A', and B'.[27] This resulted is a similar trend with the photocurrent spectroscopy in suspended monolayer $MoS_2$[28] and the heterojunction of monolayer-$MoS_2$/Si,[29] rather than with that of a photodetector based on monolayer $MoS_2$.[30]

Because Fig.2(e) strongly suggests that the role of drain voltage is more important than that of gate voltage to form p-n junctions, we disconnected the gate electrode before cooling down the ion gel and applied $V_D = 3.8$ V. We refer to this exceptional configuration as 'two terminal' because the disconnected gate electrode is floated and no longer contributes to the ion charge. In this two-terminal configuration, we still observed rectification behavior at 85 K (Fig. S7(b) in the supplementary material), which indicates the formation of p–n junctions. Therefore, we investigated the photoresponsivity of the two-terminal devices and, interestingly, we observed the 7.1 times higher photoresponsivity than that of the initial three-terminal devices at peak B as shown in Fig. 4. The improved photoresponsivity without gate biasing implies that the three-terminal configuration does not necessarily provide the ideal conditions. A key factor here may be a uniform p–n junction formation as a consequence of the self-assembly process of highly balanced anions and cations layers aligned on the $WSe_2$ surface. Further investigations, such as Kelvin probe force microscopy,[31] would be helpful to substantiate this enhancement. The sharp peaks of the excitonic transition in the spectra indicate that the photocurrent can be mainly attributed to exciton generation and subsequent dissociation, while there is relatively little contribution from free excitons. These results are compatible with the hypothesis that essentially all of the oscillator strength in TMDC monolayers is derived from the exciton transitions, and that there is no direct absorption by free carriers.[32]

In summary, we have demonstrated a monolayer $WSe_2$ EDLT-based p–n photodiode and detected the detailed excited energy dependence of its photocurrent. The two-terminal devices exhibited 7.1 times higher photoresponsivity at B resonant energy than that of the three-terminal devices, which could be because of a more uniform formation of p–n junctions as a result of the self-assembly process of highly balanced anion and cation layers on $WSe_2$ surfaces in the two-terminal devices. We observed a clear photoresponsivity peak in resonance with the excitonic absorption in the 2D crystal. Our technique enhances the understanding of the excitonic properties of p–n junctions in monolayer TMDCs, which is a crucial factor for optoelectronic devices.

**FIGURES**

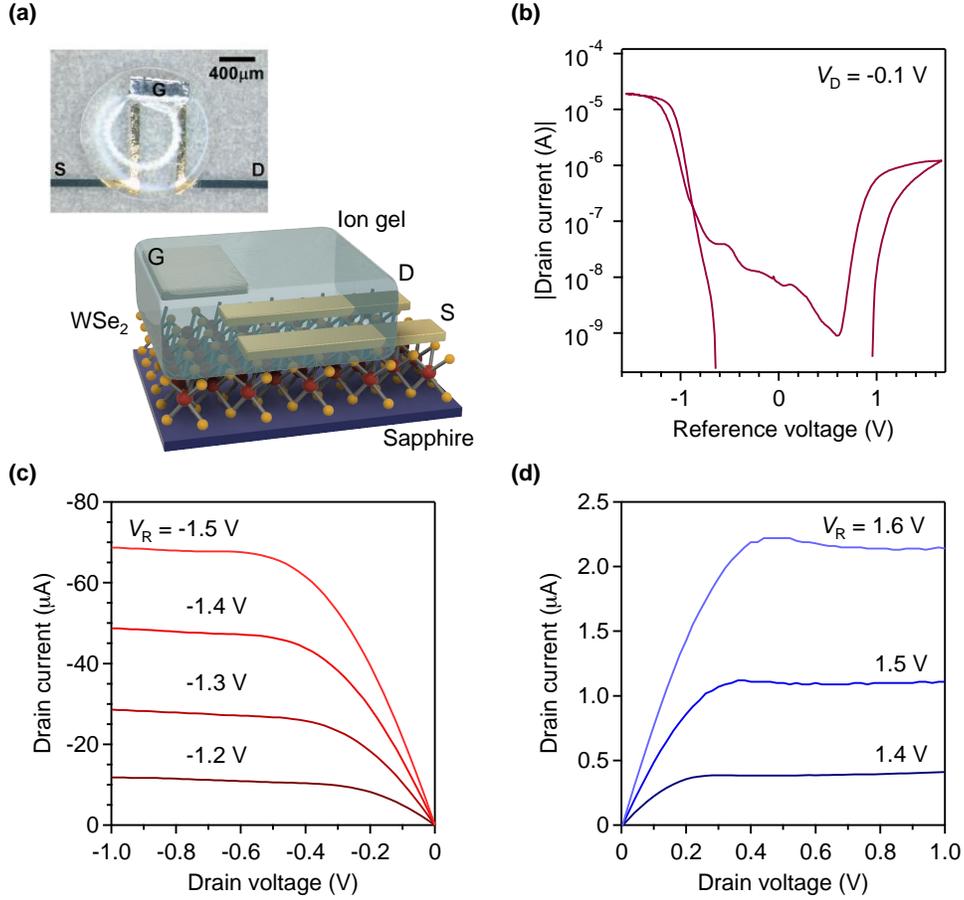

FIG. 1. (a) Picture and schematic structure of device with source (S) and drain (D) contacts of Device I. The side gate (G) electrode is also implemented for the transfer measurements. Direct formation of p–n junction on the channel of WSe$_2$ under ambipolar charge accumulation with ion gel dielectrics. (b) Transfer characteristics of the WSe$_2$ monolayer EDLT measured at the reference electrode voltage, $V_R$, where $V_R$ is the voltage of the electric double layer on the WSe$_2$ surfaces and is smaller than the applied gate voltage because the gate voltage is partially consumed by the electrical double layer on the gate electrode. The measurement is performed at room temperature, and the drain voltage, $V_D$, is fixed at –0.1 V. Output characteristics of the WSe$_2$ monolayer EDLT for (c) p- and (d) n-channel at various reference voltages.

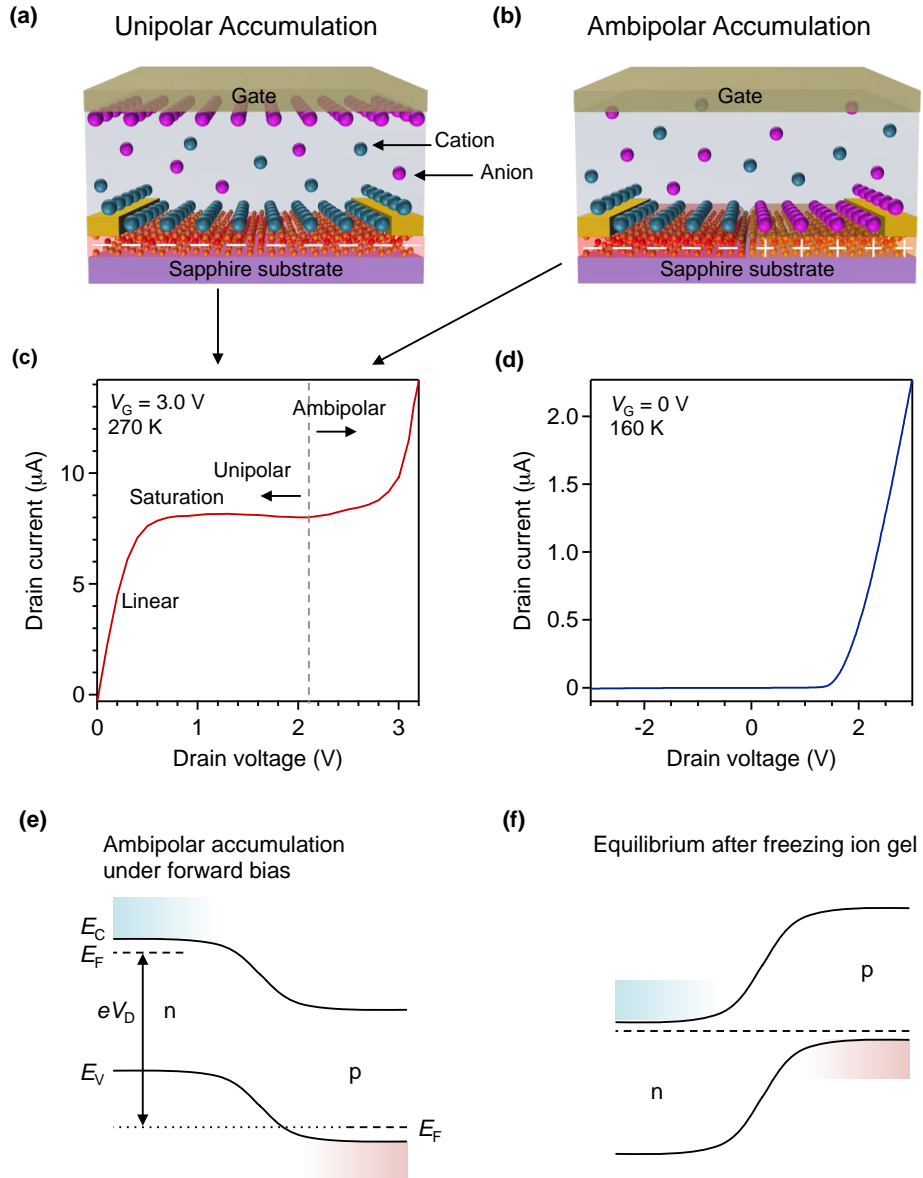

FIG. 2. (a) and (b) Schematic diagrams of unipolar and ambipolar accumulation. When the $V_D$ is markedly below the gate voltage, $V_G$, an accumulation layer is formed by one type of carrier as indicated in (a). Conversely, when $V_D - V_G$ exceeds a specific threshold voltage, the effective gate voltage at the drain electrode reverses to accumulate another type of carrier, representing ambipolar accumulation as depicted in (b). (c) Output curve of the ambipolar $WSe_2$ EDLT (Device I) measured under gate voltage of $V_G = 3.0$ V at $T = 270$ K. The evident current increase above 2.1 V following the linear and saturation indicates a simultaneous accumulation of holes and electrons. (d) Current-voltage characteristics of the p–n junction in the ambipolar $WSe_2$ EDLT at 160 K, where we froze the ion gel at $T = 160$ K to stabilize the electrostatic p–n

junctions with a constant drain/gate bias of $V_D = 3.2$ V on the signature of the electrostatic p–n junctions. Band diagram in (e) ambipolar charge accumulation under (positive $V_D$ potential) bias at $T = 270$ K and in (f) equilibrium after freezing ion gel at $T = 160$ K. $E_C$ and $E_V$ are the energy of conduction band minimum and valence band maximum, $E_F$ is the Fermi level energy (dashed lines), and $e$ is the elementary charge.

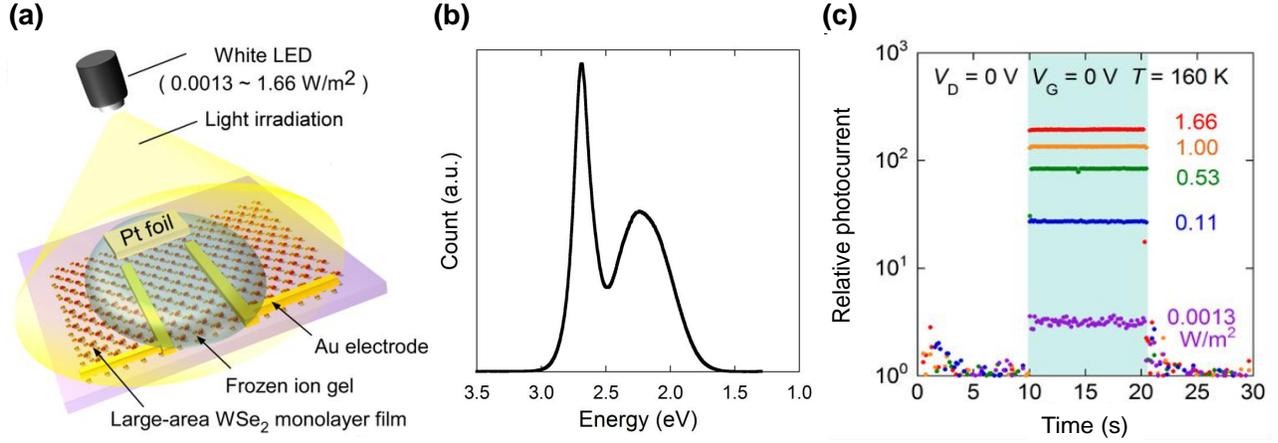

FIG. 3. (a) Schematic depiction of WSe$_2$ monolayer EDLT (Device I) with light irradiation. The ion gel covered transistor channel was irradiated with white LED light. Because the ion gels were frozen at $T = 160$ K, the p-n junction in the transistor channel was stable even without an external bias. (b) Intensity spectrum of the white LED light source. (c) Time-resolved photoresponse of the drain current measured at incident light power from 0.0013 to 1.66 W/m$^2$ on Device I. The colored areas in the chart correspond to the time of light irradiation. The photocurrent is normalized by the dark current.

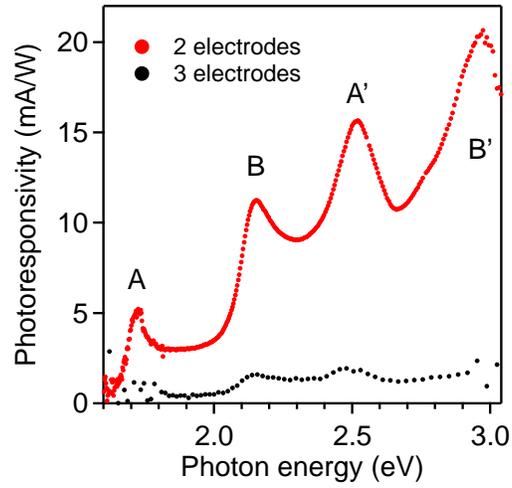

FIG. 4. Photoresponsivity spectra for two- and three-terminal devices on Device III. For the two terminal devices, the gate electrode was disconnected at room temperature with applying $V_D$ = 3.8 V, followed by cooling down the ion gel down to 85 K. For the three terminal devices, $V_D$ = 3.0 V and $V_G$ = 1.6 V were applied at room temperature, and then cooled down to 85 K. The resonance peaks, A, B, A' and B', are labeled according to earlier convention.[33]

## SUPPLEMENTARY MATERIAL

See supplementary material for the notes of photoresponse under illumination of circularly polarized light, the effect of inhomogeneity in thickness of coated ion gel, reproducibility from device to device on Device I. Absorption, Raman, PL spectra. Linear plot for the transfer characteristic on Device I. Capacitance and mobility plots. Logarithm plot for the current-voltage characteristics on Device I. Additional data set including a transfer curve, output curves under reverse and forward bias, ambipolar behavior in output curves and rectification behavior, and spectral photoresponsivity on Device II. Photocurrent under illumination of circularly polarized light on Device II. Optical image and rectification behavior in output characteristics on Device III. Intensity spectra of Xe lamp. Excitation power dependent photocurrent on Device III. Reproducibility of photoresponsivity spectra on Device III.

## ACKNOWLEDGMENTS

D.K. acknowledges the support of the Grant-in-Aid for Encouragement of Young Scientists (B) (JSPS KAKENHI Grant Number JP16K17485) from Japan Society for the Promotion of Science (JSPS). Y.W., J.P., and K.M. acknowledge the Leading Graduate Program in Science and Engineering, Waseda University, from JSPS. D.K. and J.P. were supported by the Grant-in-Aid for JSPS Fellows (JSPS KAKENHI Grant Numbers JP15J07423 and JP14J07485). T.S is grateful to the Grant-in-Aid for Encouragement of Young Scientists (A) (JSPS KAKENHI Grant Number JP26706012) from JSPS. T.T. was partially supported by the Funding Program for the Next Generation of World-Leading Researchers from JSPS, and Grants-in-Aid from MEXT (JSPS KAKENHI Grant Numbers JP16K13618, JP26102012 "π-System Figuration" and JP25000003 "Specially Promoted Research").

## REFERENCES

1. Ming-Yang Li, Yumeng Shi, Chia-Chin Cheng, Li-Syuan Lu, Yung-Chang Lin, Hao-Lin Tang, Meng-Lin Tsai, Chih-Wei Chu, Kung-Hwa Wei, Jr-Hau He, Wen-Hao Chang, Kazu Suenaga, and Lain-Jong Li, Science **349**, 524 (2015).
2. J. S. Ross, P. Klement, A. M. Jones, N. J. Ghimire, J. Yan, D. G. Mandrus, T. Taniguchi, K. Watanabe, K. Kitamura, W. Yao, D. H. Cobden, and X. Xu, Nature Nanotech. **9**, 268 (2014).
3. R. Cheng, D. Li, H. Zhou, C. Wang, A. Yin, S. Jiang, Y. Liu, Y. Chen, Y. Huang, and X. Duan, Nano Lett. **14**, 5590 (2014).

| | |
|---|---|
| 4 | B. W. Baugher, H. O. Churchill, Y. Yang, and P. Jarillo-Herrero, Nature Nanotech. **9**, 262 (2014). |
| 5 | H. Yuan, X. Liu, F. Afshinmanesh, W. Li, G. Xu, J. Sun, B. Lian, A. G. Curto, G. Ye, Y. Hikita, Z. Shen, S. C. Zhang, X. Chen, M. Brongersma, H. Y. Hwang, and Y. Cui, Nature Nanotech. **10**, 707 (2015); Jie Meng, Hua-Ding Song, Cai-Zhen Li, Yibo Jin, Lei Tang, D. Liu, Zhi-Min Liao, Faxian Xiu, and Dapeng Yu, Nanoscale **7**, 11611 (2015); D. Jariwala, V. K. Sangwan, C. C. Wu, P. L. Prabhumirashi, M. L. Geier, T. J. Marks, L. J. Lauhon, and M. C. Hersam, Proc. Natl. Acad. Sci. U. S. A. **110**, 18076 (2013). |
| 6 | A. Pospischil, M. M. Furchi, and T. Mueller, Nature Nanotech. **9**, 257 (2014). |
| 7 | V. Podzorov, M. E. Gershenson, Ch. Kloc, R. Zeis, and E. Bucher, Appl. Phys. Lett. **84**, 3301 (2004). |
| 8 | Moon Sung Kang and C Daniel Frisbie, Chem. Phys. Chem. **14**, 1547 (2013). |
| 9 | Min Sup Choi, Deshun Qu, Daeyeong Lee, Xiaochi Liu, Kenji Watanabe, Takashi Taniguchi, and Won Jong Yoo, ACS Nano **8**, 9332 (2014). |
| 10 | Yu Ye, Ziliang Ye, Majid Gharghi, Hanyu Zhu, Mervin Zhao, Yuan Wang, Xiaobo Yin, and Xiang Zhang, Appl. Phys. Lett. **104**, 193508 (2014); R. S. Sundaram, M. Engel, A. Lombardo, R. Krupke, A. C. Ferrari, P. Avouris, and M. Steiner, Nano Lett. **13**, 1416 (2013). |
| 11 | K. F. Mak, K. He, C. Lee, G. H. Lee, J. Hone, T. F. Heinz, and J. Shan, Nature Mater. **12**, 207 (2013). |
| 12 | Y Yamada, K Ueno, T Fukumura, HT Yuan, H Shimotani, Y Iwasa, L Gu, S Tsukimoto, Y Ikuhara, and M Kawasaki, Science **332**, 1065 (2011); Jiang Pu, Yohei Yomogida, Keng-Ku Liu, Lain-Jong Li, Yoshihiro Iwasa, and Taishi Takenobu, Nano Lett. **12**, 4013 (2012); L. Chu, H. Schmidt, J. Pu, S. Wang, B. Ozyilmaz, T. Takenobu, and G. Eda, Sci. Rep. **4**, 7293 (2014); Wei Xie and C Daniel Frisbie, J. Phys. Chem. C **115**, 14360 (2011); Yeonbae Lee, Colin Clement, Jack Hellerstedt, Joseph Kinney, Laura Kinnischtzke, Xiang Leng, S. D. Snyder, and A. M. Goldman, Phys. Rev. Lett. **106**, 136809 (2011). |
| 13 | Yijin Zhang, Jianting Ye, Yusuke Matsuhashi, and Yoshihiro Iwasa, Nano Lett. **12**, 1136 (2012). |
| 14 | Y. J. Zhang, T. Oka, R. Suzuki, J. T. Ye, and Y. Iwasa, Science **344**, 725 (2014). |
| 15 | Y. J. Zhang, J. T. Ye, Y. Yomogida, T. Takenobu, and Y. Iwasa, Nano Lett. **13**, 3023 (2013). |
| 16 | Daniele Braga, Ignacio Gutiérrez Lezama, Helmuth Berger, and Alberto F. Morpurgo, Nano Lett. **12**, 5218 (2012); Ismael Díez-Pérez, Zhihai Li, Shaoyin Guo, Christopher Madden, Helin Huang, Yanke Che, Xiaomei Yang, Ling Zang, and Nongjian Tao, ACS Nano **6**, 7044 (2012); Evgeniy Ponomarev, Ignacio Gutiérrez-Lezama, Nicolas Ubrig, and Alberto F Morpurgo, Nano Lett. **15**, 8289 (2015). |

17   J. T. Ye, Y. J. Zhang, R. Akashi, M. S. Bahramy, R. Arita, and Y. Iwasa, Science **338**, 1193 (2012).

18   Wenzhong Bao, Xinghan Cai, Dohun Kim, Karthik Sridhara, and Michael S Fuhrer, Appl. Phys. Lett. **102**, 042104 (2013).

19   Tobias J Octon, V Karthik Nagareddy, Saverio Russo, Monica F Craciun, and C David Wright, Adv. Opt. Mater. ASAP (2016).

20   A Abderrahmane, PJ Ko, TV Thu, S Ishizawa, T Takamura, and A Sandhu, Nanotechnology **25**, 365202 (2014); Zongyou Yin, Hai Li, Hong Li, Lin Jiang, Yumeng Shi, Yinghui Sun, Gang Lu, Qing Zhang, Xiaodong Chen, and Hua Zhang, ACS Nano **6**, 74 (2011); Wenjing Zhang, Ming-Hui Chiu, Chang-Hsiao Chen, Wei Chen, Lain-Jong Li, and Andrew Thye Shen Wee, ACS Nano **8**, 8653 (2014); H. S. Lee, S. W. Min, Y. G. Chang, M. K. Park, T. Nam, H. Kim, J. H. Kim, S. Ryu, and S. Im, Nano Lett. **12**, 3695 (2012).

21   W. Zhang, J. K. Huang, C. H. Chen, Y. H. Chang, Y. J. Cheng, and L. J. Li, Adv. Mater. **25**, 3456 (2013).

22   Tony Low, Michael Engel, Mathias Steiner, and Phaedon Avouris, Phys. Rev. B **90**, 081408(R) (2014).

23   Jing-Kai Huang, Jiang Pu, Chang-Lung Hsu, Ming-Hui Chiu, Zhen-Yu Juang, Yung-Huang Chang, Wen-Hao Chang, Yoshihiro Iwasa, Taishi Takenobu, and Lain-Jong Li, ACS Nano **8**, 923 (2013).

24   H. J. Chuang, X. Tan, N. J. Ghimire, M. M. Perera, B. Chamlagain, M. M. Cheng, J. Yan, D. Mandrus, D. Tomanek, and Z. Zhou, Nano Lett. **14**, 3594 (2014); Hongtao Yuan, Xinqiang Wang, Biao Lian, Haijun Zhang, Xianfa Fang, Bo Shen, Gang Xu, Yong Xu, Shou-Cheng Zhang, and Harold Y Hwang, Nature Nanotech. **9**, 851 (2014).

25   Kyung-Sik Shin, Hanggochnuri Jo, Hyeon-Jin Shin, Won Mook Choi, Jae-Young Choi, and Sang-Woo Kim, J. Mater. Chem. **22**, 13032 (2012); Todd G Ruskell, Richard K Workman, Dong Chen, Dror Sarid, Sarah Dahl, and Stephen Gilbert, Appl. Phys. Lett. **68**, 93 (1996).

26   Dung-Sheng Tsai, Keng-Ku Liu, Der-Hsien Lien, Meng-Lin Tsai, Chen-Fang Kang, Chin-An Lin, Lain-Jong Li, and Jr-Hau He, ACS Nano **7**, 3905 (2013).

27   L. Mattheiss, Phys. Rev. B **8**, 3719 (1973); Daichi Kozawa, Rajeev Kumar, Alexandra Carvalho, Kiran Kumar Amara, Weijie Zhao, Shunfeng Wang, Minglin Toh, Ricardo M. Ribeiro, A. H. Castro Neto, Kazunari Matsuda, and Goki Eda, Nat. Commun. **5**, 4543 (2014).

28   A. R. Klots, A. K. Newaz, B. Wang, D. Prasai, H. Krzyzanowska, J. Lin, D. Caudel, N. J. Ghimire, J. Yan, B. L. Ivanov, K. A. Velizhanin, A. Burger, D. G. Mandrus, N. H. Tolk, S. T. Pantelides, and K. I. Bolotin, Sci. Rep. **4**, 6608 (2014).

29  Bo Li, Gang Shi, Sidong Lei, Yongmin He, Weilu Gao, Yongji Gong, Gonglan Ye, Wu Zhou, Kunttal Keyshar, Ji Hao, Pei Dong, Liehui Ge, Jun Lou, Junichiro Kono, Robert Vajtai, and Pulickel M. Ajayan, Nano Lett. **15**, 5919 (2015).

30  O. Lopez-Sanchez, D. Lembke, M. Kayci, A. Radenovic, and A. Kis, Nature Nanotech. **8**, 497 (2013).

31  M Nonnenmacher, MP o'Boyle, and HK Wickramasinghe, Appl. Phys. Lett. **58**, 2921 (1991); Yang Li, Cheng-Yan Xu, and Liang Zhen, Appl. Phys. Lett. **102**, 143110 (2013).

32  M. M. Ugeda, A. J. Bradley, S. F. Shi, F. H. da Jornada, Y. Zhang, D. Y. Qiu, W. Ruan, S. K. Mo, Z. Hussain, Z. X. Shen, F. Wang, S. G. Louie, and M. F. Crommie, Nature Mater. **13**, 1091 (2014); Diana Y. Qiu, Felipe H. da Jornada, and Steven G. Louie, Phys. Rev. Lett. **111**, 216805 (2013).

33  AR Beal, JC Knights, and WY Liang, J. Phys. C: Solid State Phys. **5**, 3540 (1972).

Supplementary material

# Photodetection in p-n junctions formed by electrolyte-gated transistors of two-dimensional crystals

Daichi Kozawa[1,2,*], Jiang Pu[3], Ryo Shimizu[1], Shota Kimura[1], Ming-Hui Chiu[4], Keiichiro Matsuki[3], Yoshifumi Wada[1], Tomo Sakanoue[1,2], Yoshihiro Iwasa[5,6], Lain-Jong Li[4], Taishi Takenobu[1,2,3,7,*]

[1] Department of Applied Physics, Waseda University, Shinjuku, Tokyo 169-8555, Japan
[2] School of Engineering, Nagoya University, Furo-cho, Chikusa-ku, Nagoya, 464-8603, Japan
[3] Department of Advanced Science and Engineering, Waseda University, Shinjuku, Tokyo 169-8555, Japan
[4] Physical Science and Engineering Division, King Abdullah University of Science and Technology, Thuwal 23955-6900, Kingdom of Saudi Arabia
[5] Quantum-Phase Electronics Center and Department of Applied Physics, The University of Tokyo, Tokyo 113-8656, Japan
[6] RIKEN Center for Emergent Matter Science, Wako 351-0198, Japan
[7] Kagami Memorial Laboratory, Waseda University, Shinjuku, Tokyo 169-0051, Japan
[*] dkozawa@mit.edu
[*] takenobu@nuap.nagoya-u.ac.jp

1. **Photoresponse under illumination of circularly polarized light.**

Figure S6 is the photocurrent as a function of time, where we repeat switching circular polarization between σ⁺ and σ⁻ on Device III (a device with a channel length of 400 μm and a width of 1.2 mm which). While the work the reviewer pointed shows the circularly polarized EL, the detected photocurrent in this work shows undetectable dependence on the chirality of the incident light. As described in Ref. [1], this chirality dependence is highly related to the crystal orientation to the direction of channel or electric field. There is still room for carefully examining this chirality dependence in further sophisticated measurements.

2. **The effect of inhomogeneity in thickness of coated ion gel**

One may recognize a pattern on the surface of ion gel (Fig. S7). We should note that the small inhomogeneous thickness of coated ion gel does not fundamentally affect device performances, as electric double layer is formed only interface of metal/electrolyte and semiconductor/electrolyte. In our system, even though the thickness of the ion gel is nonuniform, the ion gel covers overall area of $WSe_2$ without patches. The electric double layer formed in ionic gel at contact with $WSe_2$ can be uniform with nanoscale thickness. This guarantees that the inhomogeneous coating of ion gel does not impact device performance.

3. **Reproducibility from device to device**

With regard to reproducibility of the series of measurements from device to device, we have tested another device with a channel length of 400 μm and a width of 1.2 mm which we call Device II. The results are shown in Fig. S5, which include (S5(a)) a transfer curve at $V_D$ = -0.1 V, output curves under (S5(b)) reverse and (S5(c)) forward bias, (S5(d)) ambipolar under $V_G$ = 3.0

V, $V_D$= 4.0 V in output curves and S5(e)) rectification behavior of device frozen at 160 K, and spectral photoresponsivity at 160 K. Hole and electron mobility are extracted from the transfer curve as 68 cm$^2$/Vs and 1.6 cm$^2$/Vs, respectively. The results that the threshold voltage for both hole and electron transport are smaller than that of the device I shown in Fig. 1(b) can be attributed that the defect density of the active material is smaller than that in device I, in which smaller gate voltage is required to achieve degenerate semiconductor to be doped. In output curve shown in Fig. S5(b) and S5(c), both biases leads to linear and saturation behaviour. We observe ambipolar accumulation (Fig. S5(d)) and rectification (S5(e)) behaviour, where rectification ratio of 38.3 is obtained. Photoresponsivity spectrum (S5(f)) follows monolayer WS$_2$ absorption spectrum. The series of the device performances is in agreement with these of Device I in good reproducibility.

For two terminal devices, the reproducibility of photoresponsivity spectrum among the measurement on the same device (Device III) is also confirmed as shown in Fig. S10. Although the Test 1 and Test 2 were conducted on different days, both are in agreement with each other. This result verifies the stability during the series of measurements.

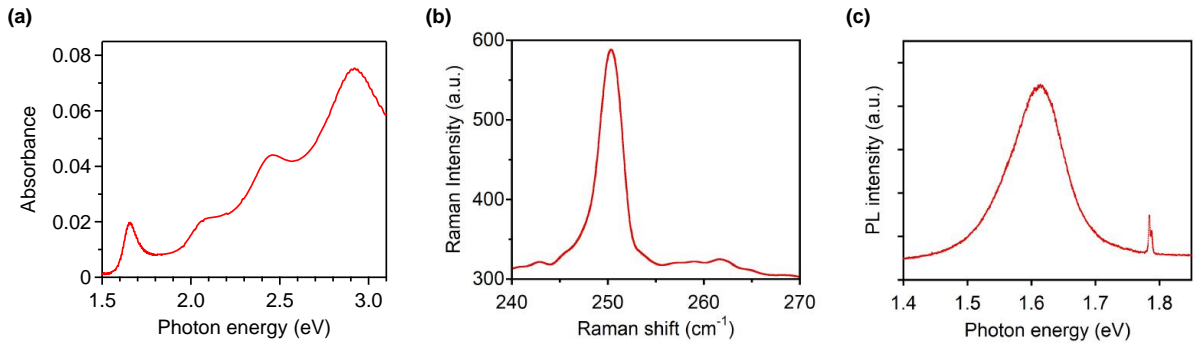

FIG. S1. (a) Absorption (b) Raman, and (c) PL spectra at room temperature.

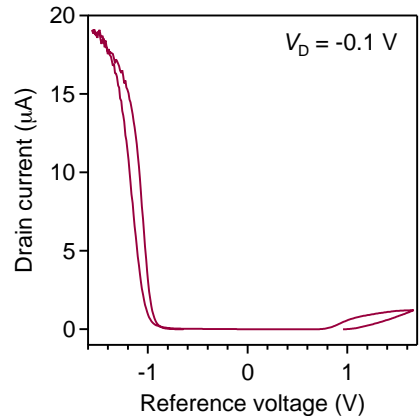

FIG. S2. Linear plot for the transfer characteristics of the WSe$_2$ monolayer EDLT of Device I.

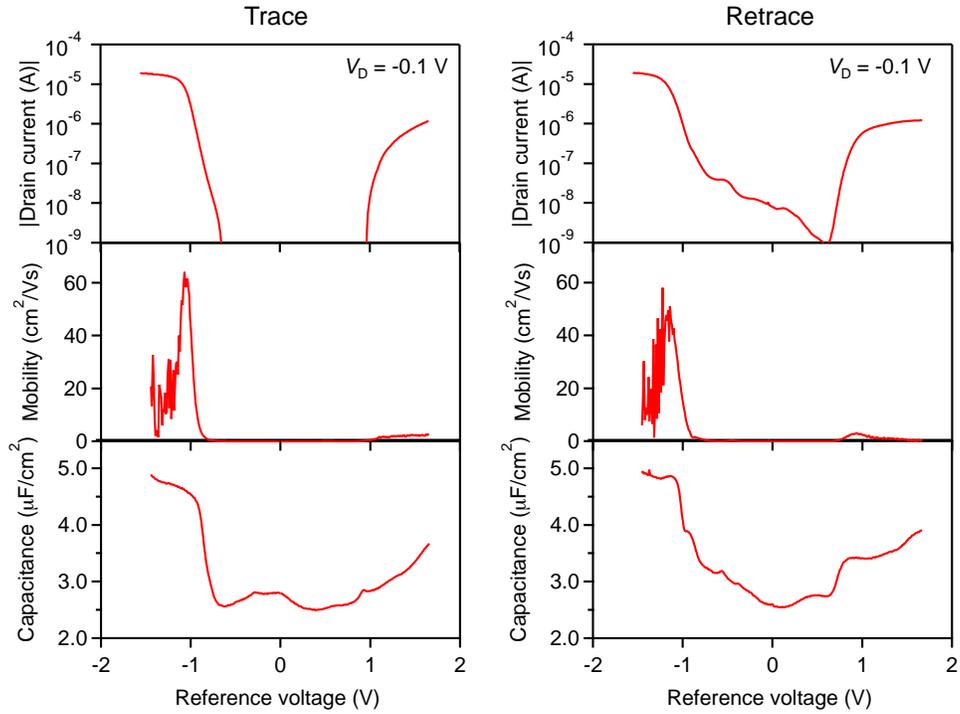

Fig. S3. Transfer (top), capacitance (middle) and mobility (bottom) curve as a function of reference voltage on Device I for trace and retrace, where these values are extracted from the transfer curve.

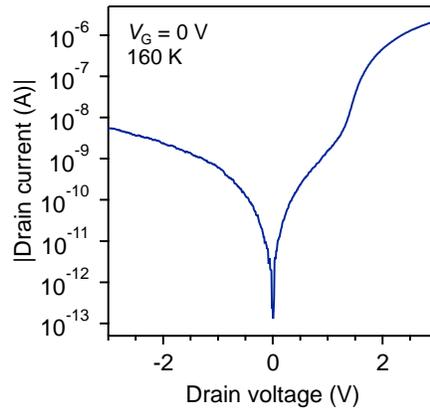

FIG. S4. Logarithm plot for the current-voltage characteristics of the p-n junction in the ambipolar $WSe_2$ EDLT of Device I at 160 K

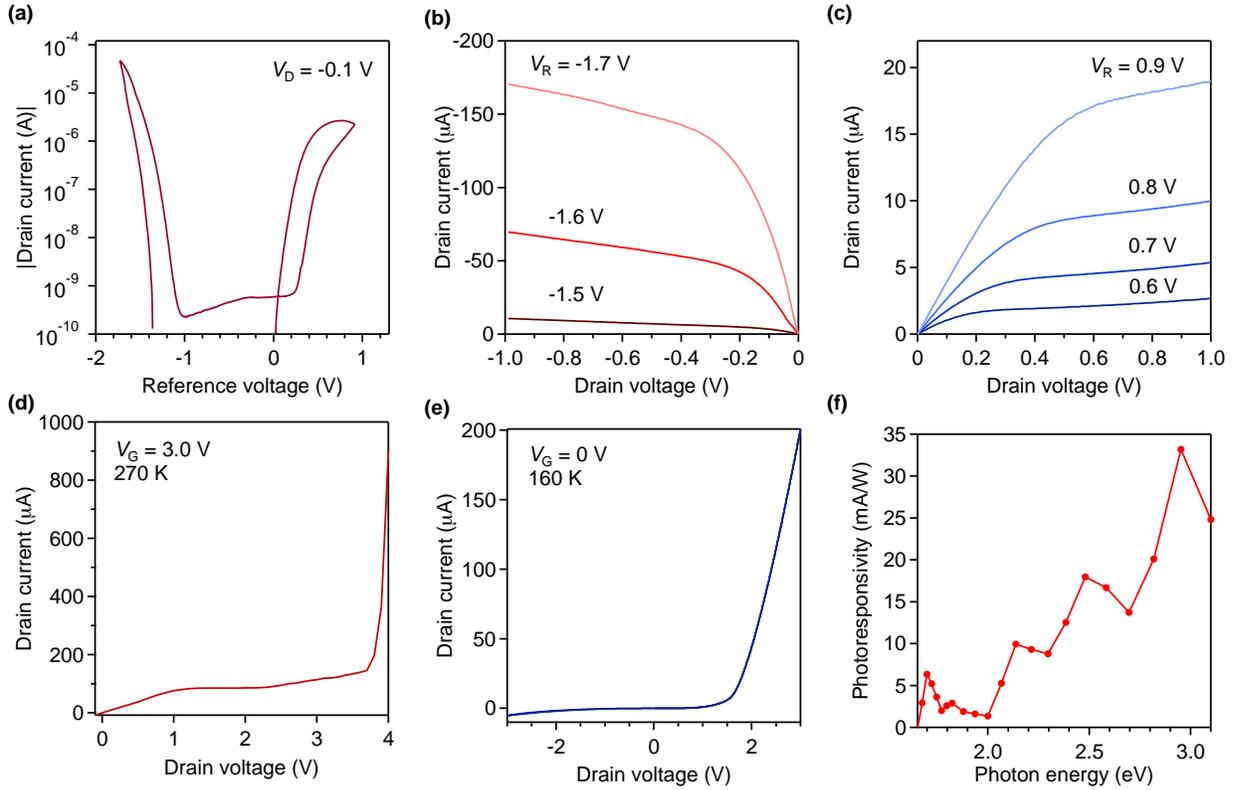

Fig. S5. Additional data set including (a) a transfer curve at $V_D = -0.1$ V, output curves under (b) reverse and (c) forward bias, (d) ambipolar under $V_G = 3.0$ V, $V_D = 4.0$ V in output curves and (e) rectification behavior of device frozen at 160 K, and spectral photoresponsivity at 160 K on Device II.

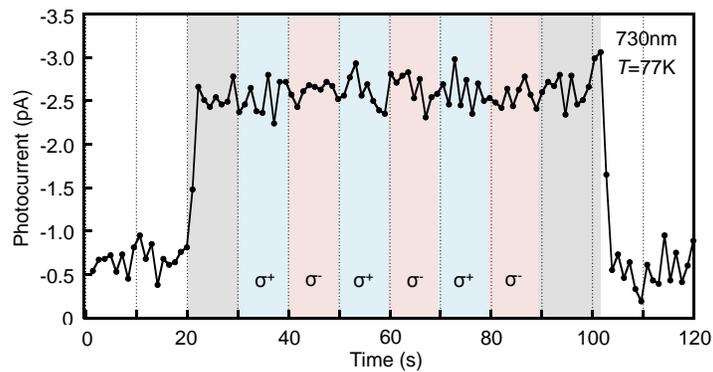

Fig. S6. Photocurrent as a function of time on Device II, where we repeat switching circular polarization between $\sigma^+$ and $\sigma^-$.

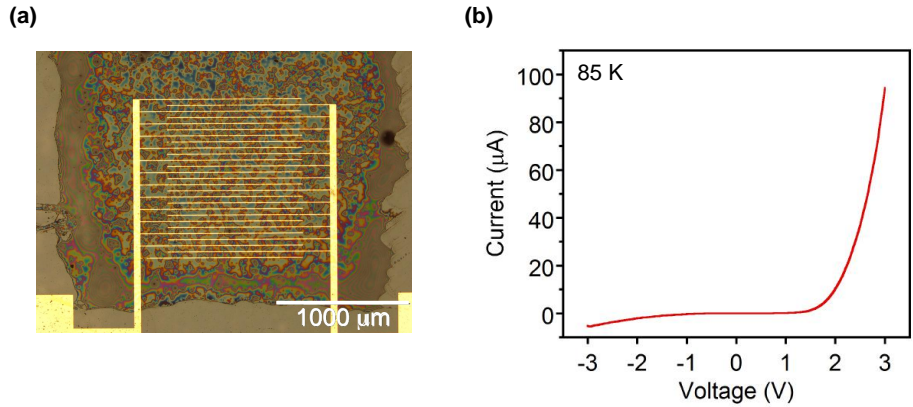

FIG. S7. (a) Optical microscope image of the Device III. The CVD-grown WSe$_2$ layer is deposited overall on the sapphire substrate. The ion gel is covered in the darker area on the device. (b) Current-voltage characteristic in formation of p-n junction at 85 K in the device shown in (a) exhibiting the diode rectification.

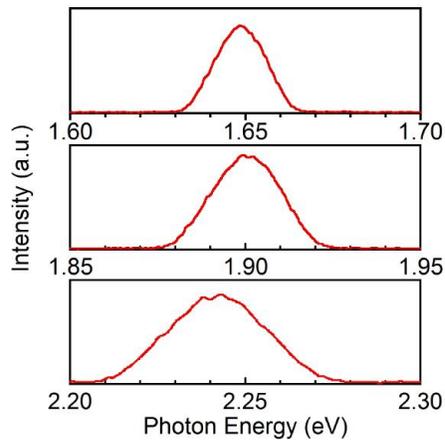

FIG. S8. Intensity spectra of Xe lamp monochromatic into various excitation energy.

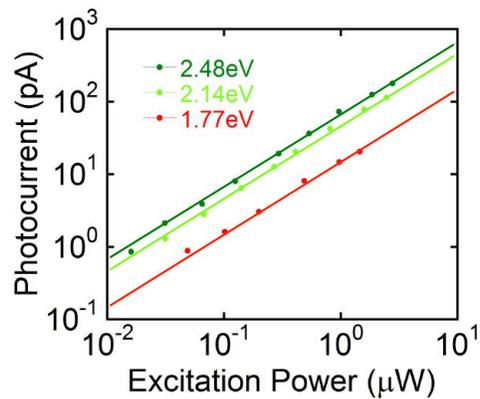

FIG. S9. Photocurrent as a function of excitation power at 1.77, 2.14 and 2.48 eV on Device III. The solid lines are fitted with linear dependence of photocurrent on excitation power.

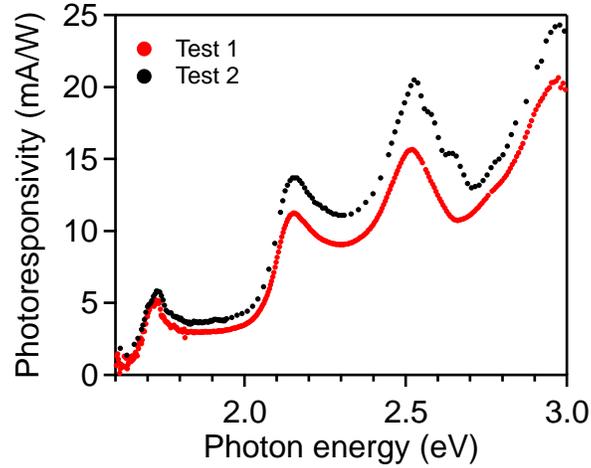

Fig. S10. Photoresponsivity spectra for two-terminal device on Device III measured at 85 K. Test 1 and Test 2 were conducted on different days The gate electrode was disconnected at room temperature with applying $V_D$ = 3.8 V, followed by cooling down the ion gel down to 85 K. $V_D$ = 3.8 V.

**REFERENCES**

[1] Y. J. Zhang, T. Oka, R. Suzuki, J. T. Ye, and Y. Iwasa, Science **344**, 725 (2014).